\documentclass[superscriptaddress,prl,twocolumn,aps,amsmath,amssymb,showpacs]{revtex4}

\usepackage{graphicx}
\usepackage{dcolumn}
\usepackage{wrapfig}
\usepackage{amsmath,amsfonts,amssymb,bm,ulem}
\usepackage[usenames]{color}

\newcommand{\ZZ}{\mathbb{Z}}
\newcommand{\Tr}{\mathop{\mathrm{Tr}}\nolimits}

\begin{document}

\date{\today}

\title{Topological multicritical point in the Toric Code
and 3D gauge Higgs Models}

\author{ I.S. Tupitsyn }
\affiliation{Pacific Institute of Theoretical Physics, University of
British Columbia, \\ 6224 Agricultural Road, Vancouver, BC V6T 1Z1, Canada}
\author {A. Kitaev }
\affiliation{California Institute of Technology, Pasadena, California 91125,
USA }
\author{N.V. Prokof'ev}
\affiliation{ Department of Physics, University of Massachusetts,
Amherst, Massachusetts 01003, USA }
\author{P.C.E. Stamp}
\affiliation{Pacific Institute of Theoretical Physics, University of
British Columbia, \\ 6224 Agricultural Road, Vancouver, BC V6T 1Z1, Canada}

\begin{abstract}
We report a new type of multicritical point that arises from
competition between the Higgs and confinement transitions in a
$\mathbb{Z}_2$ gauge system.  The phase diagram of the 3d gauge
Higgs model has been obtained by Monte-Carlo simulation on large (up
to $60^3$) lattices. We find the transition lines continue as
2nd-order until merging into a 1st-order line. These findings pose
the question of an effective field theory for a multicritical point
involving noncommuting order parameters. A similar phase diagram is
predicted for the 2-dimensional quantum toric code model with two
external fields, $h_z$ and $h_x$; this problem can be mapped onto an
anisotropic 3D gauge Higgs model.
\end{abstract}

\maketitle

\textit{Introduction.} Topological quantum phases and anyons are
well known in connection with the fractional quantum Hall effect,
but they are also expected to exist in frustrated magnets. It has
long been proposed that a certain class of resonating-valence-bond
(RVB) \cite{Anderson87} phases carries $\ZZ_2$-charges and vortices
and has a four-fold degenerate ground state on a torus
\cite{ReadChakraborty89}. A qualitative understanding of this phase
can be obtained from a so-called \textit{toric code model} (TCM)
\cite{Kitaev03}. The dimer model on the Kagome lattice is mapped
onto the TCM exactly \cite{kagome} while some other models
\cite{ReadSachdev91,MoessnerSondhi00} belong to the same
universality class.

The TCM is defined in terms of spin-$1/2$ degrees of freedom located on the
bonds of an arbitrary $2d$ lattice. The Hamiltonian is as follows:
\begin{equation}
H_{TC} = -J_x \sum_s A_s - J_z \sum_p B_p,
\label{H_TC}
\end{equation}
where $A_s = \prod_{j \in s} \sigma^{x}_j$ and $B_p = \prod_{j \in
p} \sigma^{z}_j$ are products of spin operators ($\sigma^{\alpha}_j$
are the Pauli matrixes) on the bonds incident to a lattice site $s$
and on the boundary of a plaquette $p$, respectively. The ground
state corresponds to eigenvalues $A_s=1$,\, $B_p=1$ for all $s$ and
$p$. On a surface of genus $g$, it is $4^{g}$-fold degenerate.
Elementary excitations are characterized by eigenvalues $A_s = -1$
(a $\ZZ_2$ charge on site $s$) and $B_p = -1$ (a $\ZZ_2$ vortex on
plaquette $p$); all excitations are gapped. Each type of
quasiparticle is bosonic, but due to nontrivial mutual braiding,
they must be jointly regarded as \textit{Abelian anyons}.

The Hamiltonian~(\ref{H_TC}) has special properties related to its
exact solvability: the two-point correlator vanishes and the
quasiparticles have flat dispersion. These features do not survive a
small generic perturbation, while the topological character of the
ground state and the anyonic quasiparticle statistics are robust.
Yet a sufficiently strong field can polarize the spins, driving a
transition to the topologically trivial phase. Trebst \textit{at al}
\cite{Trebst07} studied a perturbation of the form
$-h\sum_{b}\sigma^{z}_{b}$ and solved the problem by reducing it to
the $2d$ transverse-field Ising model, which is mapped to an
anisotropic $3d$ classical Ising model. In this paper we consider a
more general Hamiltonian:
\begin{equation}
H_Q = H_{TC} - h_x \sum_b \sigma^x_b - h_z \sum_b \sigma^z_b,
\label{H_2d_Q}
\end{equation}
where $b$ runs over the bonds of a square lattice and $H_{TC}$ is
given by Eq.~(\ref{H_TC}). Note that the fields $h_x$ and $h_z$
induce different types of phase transition. The term with $h_z$
creates virtual pairs of $\ZZ_2$ charges, which condense when the
field strength exceeds a certain threshold. This phenomenon may be
described as a Higgs transition, or as vortex confinement. By
duality, the field $h_x$ causes the condensation of vortices and
charge confinement. The competition of the two terms should result
in an interesting phase diagram, which is the subject of this paper.

We approach the problem by reducing the quantum Hamiltonian to a
classical \textit{anisotropic} $\ZZ_2$ gauge Higgs Hamiltonian on a
three-dimensional cubic lattice; see Eq.~(\ref{H_cl}) below. We
expect the phase diagram to be qualitatively similar to that for the
\textit{isotropic case}, i.e., model $M_{3,2}$ as defined by Wegner
\cite{Wegner71}. Monte-Carlo simulations have been performed for the
latter model because it is more amenable to numerics. Some
properties of the phase diagram in the isotropic case were predicted
by Fradkin and Shenker \cite{Fradkin79}. In particular, the
topological phase is bounded by second-order lines corresponding to
charge condensation (for $h_x \ll h_z \sim J_x, J_z$) and vortex
condensation (for $h_z \ll h_x \sim J_x, J_z$), but the two
condensate phases are continuously connected. For the quantum
Hamiltonian (\ref{H_2d_Q}), a connecting path is realized by
increasing $h_z$ so as to polarize the spins in the $z$ direction,
rotating the field in the $xz$-plane, and decreasing it again.
However, the two phase transitions are clearly different, therefore
the corresponding lines cannot join smoothly.

A previous numerical study involving $10^3$ sites by Jongeward,
Stack, and Jayaprakash \cite{Z_2_MC} showed the two lines merge into
a first-order line that partially separates the charge and vortex
condensates, and suggested the 2nd-order lines might become
1st-order before merging. As discussed below, our data for larger
systems (up to $60^3$ sites) do not confirm this conjecture. Thus,
the point where all three lines meet is likely to be a novel type of
multicritical point. Note that each of the 2nd-order transitions can
be characterized by an Ising-type order parameter, i.e., the
amplitude of the corresponding condensate. The two parameters must
somehow coexist though they are incompatible at the classical level
due to the nontrivial braiding between charges and vortices. (ie.,
the kinetic terms describing the charge and vortex transport do not
commute.)

\textit{The reduction to a classical problem.} The
Hamiltonian~(\ref{H_2d_Q}) is not gauge-invariant, but it can be
mapped to a $\ZZ_2$ gauge theory by introducing a dummy spin
variable $\mu_s$ (``matter field'') at every site, but only
considering states $|\Psi\rangle$ such that
$\mu^x_s|\Psi\rangle=|\Psi\rangle$. This constraint is a prototype
of the gauge-invariance condition,
$\mu^x_sA_s|\Psi\rangle=|\Psi\rangle$, which says that flipping the
spin on a site and all incident bonds does not change the state. To
turn one constraint into the other, we apply the transformation:
\begin{equation}
\begin{aligned}
 \sigma^z_{uv} &\to \mu^z_u\sigma^z_{uv}\mu^{z}_v,\quad
&\sigma^x_{uv} &\to \sigma^x_{uv},\\[2pt]
 \mu^z_s &\to \mu^z_s,\quad
&\mu^x_s &\to \mu^x_s A_s,
\end{aligned}
\end{equation}
where $\sigma_{uv}$ belongs to the bond connecting sites $u$ and $v$.
Then, the Hamiltonian becomes:
\begin{eqnarray}\label{gaugeH}
H = -J_{x} \sum_{s} \mu^x_{s} - J_{z} \sum_{p}B_{p}
- h_{x} \sum_{b} \sigma^{x}_{b} \nonumber \\
-h_{z} \sum_{uv} \mu^z_u \sigma^z_{uv} \mu^{z}_v.
\end{eqnarray}
Note that in the first term we have replaced $A_s$ by $\mu_s$ using the
gauge-invariance condition, $\mu^x_s A_s \equiv 1$.

We now map this 2-d quantum Hamiltonian onto a (2+1)-d classical
one. The overall scheme is standard \cite{Suzuki76}, but some care
should be taken to preserve the gauge invariance. We let
$\Delta\tau=\beta/n$, and approximate the quantum partition function
$Z=\Tr[\exp(-\beta H) {\cal P}]$ by $\Tr[\exp(-\Delta \tau \, H_{x})
{\cal P} \exp(-\Delta \tau \, H_{z})]^{n}$, where ${\cal P}$ is the
projector onto the gauge-invariant subspace and $H_x$, $H_{z}$ are
the terms in the quantum Hamiltonian that depend on
$\sigma^x_b,\mu^x_s$ and $\sigma^z_b,\mu^z_s$, respectively. This
expression can be written as a classical partition function on a
cubic lattice. The classical variables $\sigma_{b,t},\mu_{s,t}=\pm
1$, in each time slice $t$, correspond to $\sigma_{b}^{z}$ and
$\mu_{s}^{z}$ respectively. But when we change from $2d$ to $3d$,
new vertical bonds (along the time direction) appear. The classical
spins on the vertical bonds between two time slices correspond to a
choice of term in the expansion of ${\cal P} = \prod_{s} \left(
\frac{1}{2} (1+\mu^x_s A_s) \right)$. Thus we arrive at this
classical Hamiltonian:
\begin{equation}
H_{C} = - \sum_{uv} \lambda^{||, \perp}_{\mathrm{bond}}
\mu_u \sigma_{uv} \mu_v -
\sum_\mathrm{p} \lambda^{||, \perp}_{\mathrm{pl}} \prod_{j \in \mathrm{p}}
\sigma_j;
\label{H_cl}
\vspace{-0.5cm}
\end{equation}
\begin{subequations}
\begin{align}
\lambda^{||}_{\mathrm{bond}} &= - \frac{1}{2} \ln \tanh \tilde{J}_x
&-\;&\text{vertical bonds}; \label{Lambda_bv} \\
\lambda^{\perp}_{\mathrm{bond}} &= \tilde{h}_z
&-\;&\text{horizontal bonds}; \label{Lambda_bh}  \\
\lambda^{||}_{\mathrm{pl}} &= -\frac{1}{2} \ln \tanh \tilde{h}_x
&-\;&\text{vertical plaquettes}; \label{Lambda_pv} \\
\lambda^{\perp}_{\mathrm{pl}} &= \tilde{J}_z
&-\;&\text{horizontal plaquettes}, \label{Lambda_ph}
\end{align}
\end{subequations}
where $\tilde{J}_x=J_x\Delta\tau$,\, $\tilde{J}_z=J_z\Delta\tau$,\,
$\tilde{h}_x=h_x\Delta\tau$,\, $\tilde{h}_z=h_z\Delta\tau$. This
model is an anisotropic generalization of the $\ZZ_2$ gauge Higgs
model \cite{Fradkin79}.

As a final step, we eliminate the redundancy by fixing $\mu_{s}$. This only
changes the classical partition function by a constant factor since
Hamiltonian~(\ref{H_cl}) can be written in terms of the gauge-invariant
variables $S_{uv}=\mu_u\sigma_{uv}\mu_v$:
\begin{equation}\label{H_cl1}
\widetilde{H}_{C} = - \sum_{b} \lambda^{||, \perp}_{\mathrm{bond}}
S_{b} -
\sum_{\mathrm{p}} \lambda^{||, \perp}_{\mathrm{pl}} \prod_{j \in \mathrm{p}}
S_j.
\end{equation}
More detailed calculations show that the quantum and classical
partition functions are related by
\begin{equation}\label{Z_Q_C}
Z=\left(\tfrac{1}{2}\sinh(2\tilde{J}_x)\right)^{k/2}
\left(\tfrac{1}{2}\sinh(2\tilde{h}_x)\right)^{m/2}\,\widetilde{Z}_{C},
\end{equation}
where $k$ and $m$ are the number of vertical bonds and plaquettes,
respectively.

Of course, Eq.~(\ref{Z_Q_C}) holds only in the limit $\Delta\tau\to
0$. However, we take the liberty of parametrizing the general
classical Hamiltonian~(\ref{H_cl1}) by $\tilde{J}_x$, $\tilde{J}_z$,
$\tilde{h}_x$, $\tilde{h}_z$, even though the corresponding quantum
problem may not be defined. In the isotropic case, two parameters
will suffice:
\begin{equation}
\widetilde{H}_C = - \lambda_{\mathrm{bond}} \sum_{b} S_b -
\lambda_{\mathrm{pl}} \sum_{\mathrm{p}} \prod_{j \in \mathrm{p}} S_j,
\label{H_C_S}
\vspace{-3mm}
\end{equation}
where $\lambda_{\mathrm{bond}} = \tilde{h}_z$, $\lambda_{\mathrm{pl}} =
-\frac{1}{2} \ln \tanh \tilde{h}_x$. This model is equivalent to the
isotropic $\ZZ_2$ gauge Higgs model \cite{Fradkin79} and in what follows
we compute its phase diagram.

\textit{Phase diagram in the isotropic case.} At
$\lambda_{\mathrm{bond}} = 0$ all configurations, including ground
states, have the same degeneracy factor $2^{2N}$. The actual
physical variables in this limit are plaquette numbers $ N_p =
\prod_{j \in \mathrm{p}} S_j$, and the model itself is \textit{dual}
to the 3D classical Ising model (Eq.~(\ref{H_C_S}) is also known as
the 3D Ising gauge theory \cite{Wegner71}). Using high-accuracy
results of Ref.~\cite{Gupta96} for the critical point and the
duality relation $\lambda_{\mathrm{pl}} = -1/2 \ln \tanh (J/T)$,
where $J$ is the Ising exchange coupling, we obtain
$\lambda^{(c)}_{\mathrm{pl}} = 0.7614125$.

At arbitrary values of $\lambda_{\mathrm{bond}}$ and
$\lambda_{\mathrm{pl}}$ the model is \textit{self-dual}
\cite{Balian75}, i.e. it maps to itself under the coupling constant
transformation $\lambda_{\mathrm{bond},\mathrm{pl}} \to -1/2 \ln
\tanh(\lambda_{\mathrm{pl},\mathrm{bond}})$. This means that the
phase diagram has a symmetry, or \textit{self-duality}, line defined
by $\lambda_{\mathrm{bond}} = -1/2 \ln
\tanh(\lambda_{\mathrm{pl}})$. Under the duality mapping
$(\lambda_{\mathrm{bond}}=0, \lambda_{\mathrm{pl}}=0.7614125) \to
(\lambda_{\mathrm{bond}} = 0.221655, \lambda_{\mathrm{pl}}=\infty)$,
which gives us two Ising-type critical points on the phase diagram.

To calculate the rest of the phase diagram we performed Monte Carlo
simulations using standard single-spin flip updates, supplemented by
rare (once per $N^2$ updates) flips of all spins belonging to bonds
cut by planes oriented along any one of the crystal axes, or along
any of the diagonals to these axes. There are $9 N$ possible planes
satisfying this condition, and we select any of them at random. The
plaquette energy (second term in (\ref{H_C_S})) is conserved by this
update. To determine the 2nd-order critical lines, we employed a
standard finite-size scaling analysis of the specific heat $C_v$,
for linear system sizes $N=24, \; 36, \; 48,$ and $60$ (ie., for $3
N^3$ spins). First-order critical points were identified and located
using energy distributions. These distributions are bi-modal (have
two maxima) for the first-order transitions and single-modal
otherwise. We thermalized our samples for up to $10^6$ MC sweeps
(one sweep having $3 N^3$ elementary updates). The data were
accumulated for $\sim 4 \times 10^8$ MC sweeps.

\begin{figure}[h]
\vspace{-2.9cm}
\includegraphics[width=7.6cm]{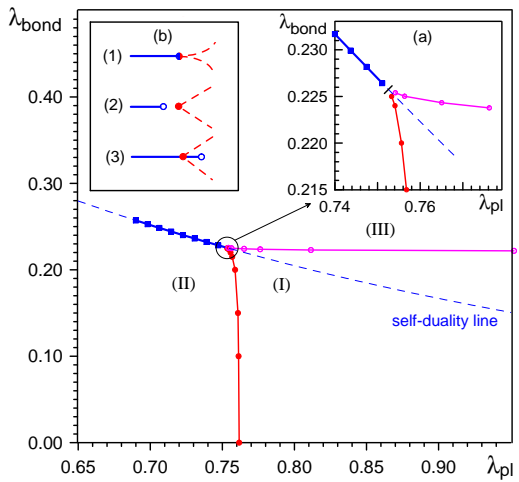}
\vspace{-2.2cm}
\caption{ (color online). The phase diagram of the
Hamiltonian (\ref{H_C_S}). Circles correspond to the second-order
transitions (open and filled symbols are related by the duality
transformation). Filled squares describe the first-order self-dual
transitions. Bold and dashed lines are used to guide an eye and
correspond to the 1st- and 2nd-order transitions, respectively. The
phases are: (I) - topological phase; (II) - topologically disordered
phase; (III) - magnetically ordered phase. In inset (a) we show the
region where all phases meet each other. In inset (b) we show three
alternative ways of connecting the lines.}
\label{Fig1}
\end{figure}

\begin{figure}[h]
\vspace{-2.8cm}
\includegraphics[width=8cm]{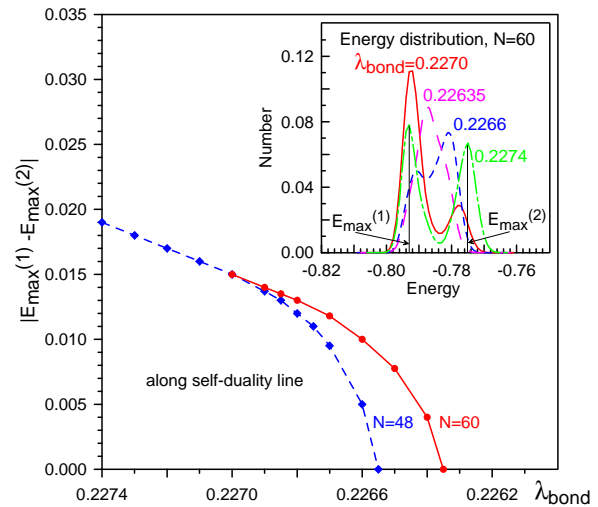}
\vspace{-2.4cm} \caption{ (color online). The distance between two
maxima in bi-modal energy distributions along the self-duality line
for $N=60$ as a function of $\lambda_{\mathrm{bond}}$. The inset
shows examples of the energy distributions at various values of
$\lambda_{\mathrm{bond}}$.}
\label{Fig2}
\end{figure}

The resulting phase diagram is presented in Fig.\ref{Fig1}. The
first-order transition coinciding with the self-duality line was
observed for $0.2575(5)
> \lambda_{\mathrm{bond}} > 0.22635(5)$. Outside of this interval we
saw no bi-modal structure in the energy distribution for system
sizes up to $N=60$. The inset of  Fig.\ref{Fig2} shows the evolution
of the energy distribution function along the self-dual line. Even
when the bi-modal structure is observed it is extremely weak,
developing only for large $N$, and the distribution can be sampled
in the minimum without flat-histogram or similar reweighting
techniques.

As noted above, these results conflict with previous MC simulations
in Ref.~\cite{Z_2_MC}, who suggest the 1st-order line splits into
two 1st-order lines. The inset (a) of Fig.\ref{Fig1} shows a closeup
of the controversial region. Though we were able to resolve critical
points with an accuracy of at least three digits, we observed no
splitting of the self-dual 1st-order line into two 1st-order
transitions. We also find no evidence for tri-critical points on the
Ising-type lines as long as we can resolve two separate transitions.
There remains a tiny parameter range between the apparent
disappearance of the bi-modal distribution on the self-dual line
(this disappearance probably due to our limited system size) and two
resolved 2nd-order transitions.

To probe the behavior in this tiny parameter range we need a
different approach. We therefore scanned energy distributions at
$30$ points ($N=48$) along the line perpendicular to the
self-duality line right in the questionable region (short solid line
in the inset (a)). If the 1st-order line were to split above the
scan, the third maximum would have to emerge in the energy
distribution right between the two maxima we observe on the
self-dual line - implying that the energy maxima on the self-dual
line could not merge smoothly, and right below the split, three
maxima would have to be seen in the energy distribution. However all
distributions along the scan were found to have only one peak. It is
also clear from the main part of Fig.\ref{Fig2} that on the
self-duality line, the energy maxima approach each other and merge
continuously as $\lambda_{\mathrm{bond}}$ increases. The curves
presented in Fig.\ref{Fig2} follow a power law near the vanishing
point, with corresponding critical exponent $\sim 0.55$.

We thus conclude that the split 1st-order scenario does not work.
Instead there are three possibilities. Either all three lines merge
at one point (case $(1)$ in the inset (b), Fig.\ref{Fig1}); or the
1st-order line ends \textit{before} or \textit{after} the point
where two 2nd-order lines touch the self-dual line (cases $(2)$ and
$(3)$ in the inset (b), Fig.\ref{Fig1}). Unfortunately our data
cannot distinguish between the alternatives because the 2nd-order
lines seem to touch at extremely small (possibly zero) angle.
Formally, option (2) fits the data best. Theoretically, the last two
scenarios are less demanding since they fit the existing theory of
phase transitions (our data suggest that the 2nd-order transitions
cannot merge into a single smooth curve and form a kink at the
self-dual line). We are not aware of any effective theory leading to
the first scenario.

\begin{figure}[h]
\vspace{-2.6cm}
\includegraphics[width=8cm]{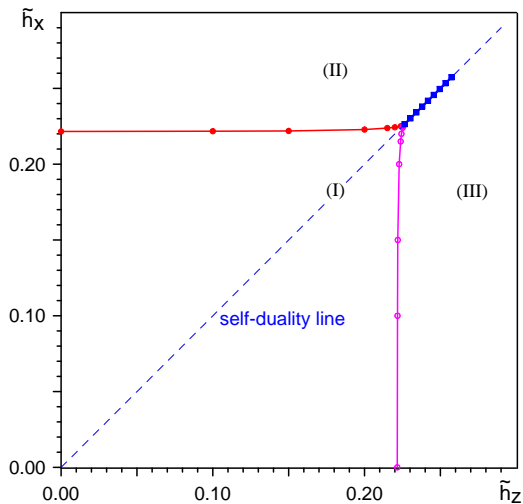}
\vspace{-2.6cm}
\caption{ (color online). The phase diagram Fig.\ref{Fig1} in terms of the
renormalized external fields, $\tilde{h}_x = h_x \Delta \tau$ and $\tilde{h}_z
= h_z \Delta \tau$, of the model (\ref{H_2d_Q}). The phases (I), (II) and (II)
are the same as in Fig.\ref{Fig1}.}
\label{Fig3}
\end{figure}

\textit{Phases.} Using the two correspondence equations
\begin{subequations}
\begin{eqnarray}
\tilde{h}_z = - \frac{1}{2} \ln \tanh (\tilde{J}_x) =
\lambda_{\mathrm{bond}}; \label{h_z_tau} \\
\tilde{J}_z = - \frac{1}{2} \ln \tanh (\tilde{h}_x) =
\lambda_{\mathrm{pl}}, \label{J_z_tau}
\end{eqnarray}
\end{subequations}
we can reformulate the phase diagram Fig.\ref{Fig1} in terms of the renormalized
parameters $\tilde{J}_x$, $\tilde{J}_z$, $\tilde{h}_x$ and $\tilde{h}_z$ of the
TCM. The resulting phase diagram in terms of the external fields is presented
in Fig.\ref{Fig3}. Let us go through the phases in this Figure.

The phase (I) corresponds to the topological phase of the model
(\ref{H_2d_Q}) (the ``free charge'' phase of the isotropic $\ZZ_2$
Higgs model (HM), \cite{Fradkin79}). In this phase the system tends
to have all $B_p = 1$ and $A_s = 1$ and a realization of such a
state is obviously not unique. The plaquettes with $B_p = -1$
(magnetic vortices) and vertices with $A_s = -1$ (electric charges)
appear mainly in the vicinity of the critical lines between the
phases (I) and (II) (vortices) and (I) and (III) (charges). The
phase (III) may be called ``magnetically ordered'' since the spins
are mostly polarized in the $z$-direction. However, $\langle
\sigma^z \rangle$ also has nonzero value everywhere in the phase
diagram. The true order parameter may be written as
$\langle \mu^z \rangle$ using the gauge-symmetrized Hamiltonian
(\ref{gaugeH}). A non-zero value of this parameter results in the
confinement of magnetic vortices (no free vortices) and the condensation
of electric charges. In the HM this is the ``Higgs'' phase.  The phase
(II) is characterized by a dual order parameter related to
$\langle \sigma^x \rangle$, which can be defined by rewriting the
Hamiltonian in different variables. Its nonzero value results in
non-conservation of total magnetic ``charge'' and condensation of
magnetic vortices while electric charges are confined (no free
charges). This phase corresponds to the ``confinement'' phase of the
HM. The transition between the phases (II) and (III) is accompanied
by a sharp change in the number of vortices and charges,
corresponding to a ``liquid-gas'' type transition. The self-duality
symmetry reflects the symmetry between charges and vortices.

\textit{Summary.} The topological phase of the toric code model (the
``free charge'' phase of the 3d gauge Higgs model) remains stable in
a rather wide range of fields and breaks down via two Ising type
transitions whose critical lines meet with the 1st-order one
corresponding to a liquid-gas type transition. The 1st-order line
either meets with two 2nd-order lines in one multicritical point, or
terminates before or after the point where two 2nd-order lines touch
the self-duality line. The construction of an effective field theory
for this multicritical region is an interesting open problem.

We thank E. Fradkin, B. Svistunov, S. Trebst, M. Troyer, I. Affleck, and
K. Shtengel for discussions. We are also indebted to M. Berciu and J. Heyl
whose research clusters were used to perform our MC simulations.

\end{document}